\documentclass[prl,aps,preprintnumbers,twocolumn,amsmath,amssymb,superscriptaddress]{revtex4}
\usepackage{graphicx}
\usepackage{dcolumn}
\usepackage{bm}
\begin{document}
\preprint{\textit{Version}\date{\today}}
\title{Electrical control of the exciton-biexciton splitting in a single self-assembled InGaAs quantum dots}
\author{M. Kaniber}
\email{kaniber@wsi.tum.de}
\author{M. F. Huck}
\author{K. M\"{u}ller}
\author{E. C. Clark}
\affiliation{Walter Schottky Institut, Technische Universit\"at M\"unchen, Am Coulombwall 4, 85748 Garching, Germany}
\author{F. Troiani}
\affiliation{INFM \& Physics Department, University of Modena and Reggio Emilia, Via Campi 213A, 41100 Modena, Italy}
\author{M. Bichler}
\affiliation{Walter Schottky Institut, Technische Universit\"at M\"unchen, Am Coulombwall 4, 85748 Garching, Germany}
\author{H. J. Krenner}
\affiliation{Institut f\"ur Physik, Lehrstuhl f\"ur Experimentalphysik I, Universit\"at Augsburg, 86150 Augsburg, Germany}
\author{J. J. Finley}
\affiliation{Walter Schottky Institut, Technische Universit\"at M\"unchen, Am Coulombwall 4, 85748 Garching, Germany}
\date{\today}
\begin{abstract}
We report on single InGaAs quantum dots embedded in a lateral electric field device. By applying a voltage we tune the neutral exciton transition into resonance with the biexciton using the quantum confined Stark effect. The results are compared to theoretical calculations of the relative energies of exciton and biexciton. Cascaded decay from the manifold of single exciton-biexciton states has been predicted to be a new concept to generate entangled photon pairs on demand without the need to suppress the fine structures splitting of the neutral exciton. 
\end{abstract}
%
\maketitle
%
%
The controlled next generation of entanglement is an important concept in both quantum information science \cite{Bouwmeester97} and quantum cryptography \cite{Ekert91}. Benson \textit{et al.} \cite{Benson00} proposed the use of biexciton-exciton cascade in semiconductor quantum dots (QDs) to generate such nonclassical states of light. Subsequently, many groups \cite{Stevenson06, Akopian06, Hafenbrak07} worldwide have demonstrated the generation of entangled photon pairs via this process. Creating polarization entangled, rather than classically correlated \cite{Santori02}, photon pairs has been achieved by tuning the exciton fine structure splitting to zero \cite{Bayer99}. Recently, this has been achieved by several groups \cite{Gerardot07, Vogel07, Kowalik07} by applying an electric field in a lateral geometry in the base plane of the QD. An alternative approach for realizing an entangled photon source, proposed by Avron \textit{et al.} \cite{Avron08}, requires the exciton to be tuned into resonance with the biexciton in order to exploit time reordering of the emitted photon pairs.\\ 
%
%
In this Letter we demonstrate tuning of exciton and biexciton states of a single InGaAs QD into resonance by applying an electric field parallel to the QD layer. Comparison with theory shows that the sign of the energy shift arising from the quantum confined Stark effect \cite{Fry00} is opposing for the exciton and biexciton states. This allows us to bring both transitions into resonance at electric fields of F$\sim$17 kV/cm. Such devices are promising candidates for realizing an electrically tunable source of entangled photon pairs.\\
%
%
\begin{figure}[t]
    \begin{center} 
       \includegraphics[width=0.8\columnwidth]{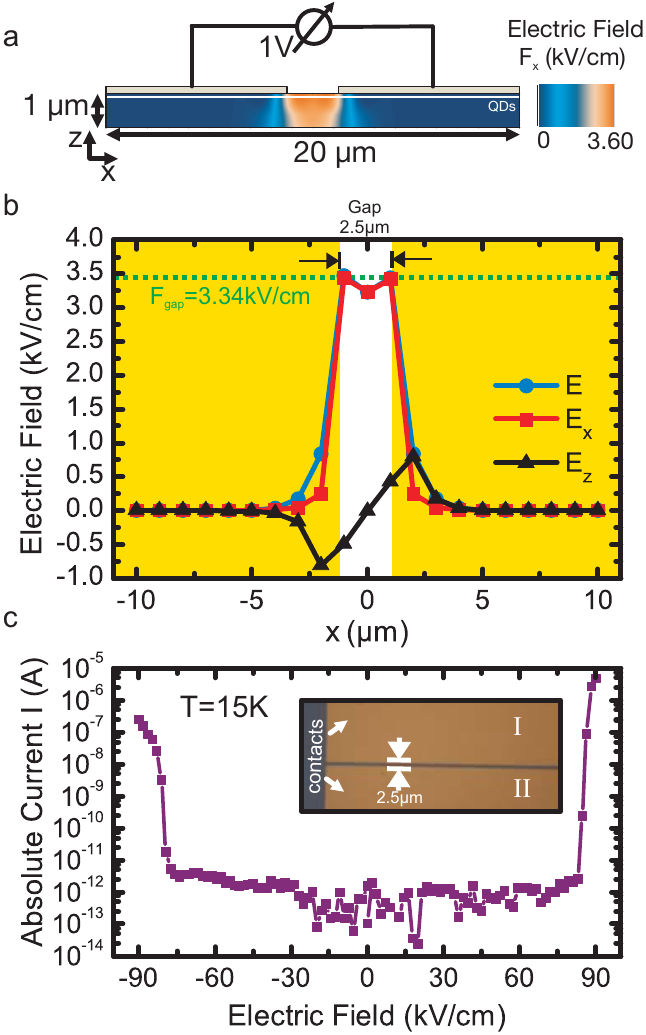}
    \end{center}    
    \caption{ (color online) (a) Simulated electric field distribution for the split-gate geometry for an applied voltage of $1$~V. (b) Simulated electric field at the QDs ($100$~nm below 
     the sample surface) for the x-component (squares), z-component (triangles) and the total electric field (circles).(d) Current as a function of applied electric field for a back-to-back Schottky 
     diode at T=15 K without illumination. (inset) Top view photograph of a completely processed sample with a $2.5\mu$m gap wide gap.}
\end{figure}
The samples investigated were grown by molecular beam epitaxy and consist of the following layer sequence: starting with a semi-insulating (100) GaAs wafer we deposited a $300$~nm thick GaAs buffer layer, followed by 25 periods of a AlAs ($2.5$~nm)/GaAs ($2.5$~nm) superlattice. We grew a $200$~nm thick GaAs waveguide, into the center of which a single layer of nominally In$_{0.5}$Ga$_{0.5}$As QDs was incorporated. The areal density of the QDs was varied during the growth by stopping the rotation of the wafer and samples with a density of $\sim$10 $\mu$m$^{-2}$, as revealed by atomic force microscope measurements (not shown here), were chosen for further processing. Using a combination of optical lithography and electron beam metalization we established back-to-back Ti/Au Schottky gates on the surface of the sample. In a first evaporation step we deposit a $20$~nm titanium undercoating, followed by a 60 nm~gold layer. In a second lithography step we produced larger Ti/Au bonding pads with a layer thickness of 20 nm~and 200 nm, respectively. In the inset of Fig. 1(c) we present a top view photograph of a fabricated device showing the contacts, labeled I and II, separated by a 2.5 $\mu$m wide gap. We simulated the electric field distribution between I and II by using the finite element analysis package QuickField \cite{Quick} which solves a two-dimensional Poisson equation for a given geometry with certain boundary conditions as the dielectric constant for GaAs ($\epsilon_{GaAS}=13$) and the electrostatic potential difference between the two diode constants. The result of the simulation for our contact geometry with an applied potential set to $1$~V is shown in Fig. 1(a). The electric field is encoded in color and the QD layer is indicated by the white line, approximately $100$~nm below the sample surface. The electric field within the gap between the contacts and in the region of the QD is homogeneous and almost purely lateral as shown in Fig. 1(b) where we plot the electric field in x- and z-direction as a function of the distance from the gap center. From this simulation we obtain a conversion factor of $F_{gap}=3.34$~kV/cm for an applied voltage of $1$~V. A typical current characteristic recorded at $T=15$~K without illumination is shown in Fig. 1(c). This clearly shows that the devices fabricated have very low leakage currents $<1$~pA, close to the resolution limit of the measurement device, for electric fields in the interval  $-75$~kV/cm to $+75$~kV/cm. This indicates excellent current blocking of the back-to-back Schottky diodes, allowing us to apply lateral electric fields from $-75$~kV/cm to $+75$~kV/cm to our dots without perturbing currents flowing.\\
%
%
The optical studies presented below were performed using a confocal micro-photoluminescence ($\mu$-PL) setup that provides a spatial resolution of $\sim700$~nm at emission wavelength $950$~nm. The sample was mounted in a helium-flow cryostat to achieve $T=15$~K and electrically connected to a Keithley 2601 source meter to apply the bias voltage. The sample was excited quasi-resonantly, into the wetting layer (E$_{ex}^{WL}=1.46$~eV) using a continuous wave Ti:sapphire laser. The PL signal of single QDs was collected using an $100\times$ microscope objective (numerical aperture=0.8), spectrally dispersed by a $0.55$~m imaging monochromator and detected with a Si-based, liquid nitrogen cooled charge coupled device detector.\\ 
%
\begin{figure}[t]
    \begin{center} 
       \includegraphics[width=0.8\columnwidth]{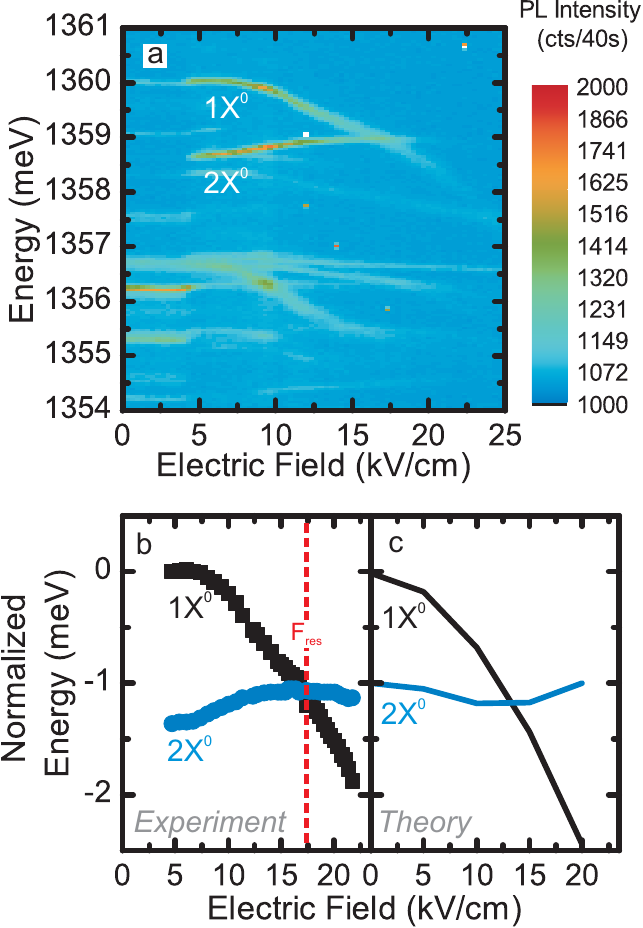}
    \end{center}
    \caption{ (color online) (a) PL spectra recorded as a function of electric field from $0$~kV/cm to $25$~kV/cm (false color plot). (b) Extracted normalized energy of the emission lines 1X$^0$ and 2X$^0$ as a function of electric field resulting in a resonance at $F_{res}\sim17$~kV/cm. (c) Calculations of the renormalization energies of the 1X$^0$ and 2X$^0$ lines of a single QD in a lateral electric field geometry.}
\end{figure}
%
%
In Fig. 2(a) we present typical $\mu$-PL measurements of a single QD as a function of applied electric field in the interval $0$~kV/cm to $+25$~kV/cm in a false color plot. Two dominating emission peaks are observed for electric fields F$>4.5$~kV/cm, labeled 1X$^0$ (E$_{1X^0}$=1360.02 meV) and 2X$^0$ (E$_{2X^0}$=1358.66 meV), respectively. These lines are attributed to the single neutral exciton and biexciton, respectively (see discussion below). For electric fields in the range $0$~kV/cm$<$F$<4.5$~kV/cm, the spectrum is dominated by numerous other emission lines that are shifted in energy and abruptly quench when the critical field, F$_{crit}=4.5$~kV/cm, is reached. As discussed below, we attribute this behavior to the appearance of charged states in the QD due to field induced ionization of excitons pumped into the wetting layer. This gives rise to many emission lines due to the enhanced number of final states during the spontaneous emission process due to the statistical capture of electrons and holes into the QD. We will discuss this idea in more detail below.\\
For electric fields F$>4.5$~kV/cm the emission energy of 1X$^0$ decreases whilst the emission energy of 2X$^0$ increases. This behavior is attributed to the quantum confined Stark effect \cite{Fry00}. The strongly different DC Stark shift of the 2X$^0$ line compared to 1X$^0$ results primarily from the repulsive Coulomb interaction between the two holes or/and the two electrons in the biexciton that are not precisely compensated by the attractive e-h interaction. This effect appears predominantly for lateral electric fields since the electric field is applied parallel to the base of the QDs which is typically $\sim30$~nm and, therefore, much larger than its height of typically $\sim5$~nm \cite{Krenner05}. The opposing DC Stark shift allows the 1X$^0$ and 2X$^0$ lines to be tuned into resonance at an electric field F$_{res}\sim17$~kV/cm, as shown in Fig. 2(b). Further increase of the electric fields tunes the 1X$^0$ and 2X$^0$ out of resonance once again. At large applied fields we observe a pronounced decrease of the PL intensity due to a combination of suppressed charge carrier capture into the QDs from the wetting layer and enhanced carrier tunneling escape out of the QD.\\
To support these ideas we calculated the DC Stark shift of the 1X$^0$ and 2X$^0$ states in a QD. The calculations were performed by considering a parabolic in-plane potential and including Coulomb interactions. In Fig. 2(c) we present calculations of the renormalization energies of the 1X$^0$ and 2X$^0$ lines for a disk-shaped InGaAs/GaAs QD (base length $18$~nm and height $4$~nm) as a function of electric field. In order to calculate the transition energies of the two states we considered the initial and final state of those transitions. For the 1X$^0$ state the final state is an empty QD. Therefore, the observed Stark shift is completely determined by the field induced modification of the Coulomb interaction between the electron and hole, V$^{eh}$. This results in a quadratic Stark shift towards lower energy according to \cite{Fry00}
\begin{equation}
  E=E_0-\vec{p}_0\cdot\vec{F}-\beta_{1X^0}\cdot\vec{F}^2
\label{eqn1}
\end{equation}
where E$_0$, $\vec{p}_0$, $\beta_{1X^0}$ and $\vec{F}$ denote the exciton energy at zero electric field, the intrinsic dipole moment at zero electric field, the polarizability of the electron-hole pair and the applied electric field, respectively.\\
To a first approximation, the transition energy of the 2X$^0$ state as a function of electric field is given by
\begin{equation}
  \Delta E_{2X^0}(F)=E_{2X^0}^0-(\beta_{2X^0}-\beta_{1X^0})\cdot\vec{F}^2
\label{eqn2}
\end{equation}
where E$_{2X^0}^0$ and $\beta_{2X^0}$ denote the biexciton energy at zero electric field and the polarizability of the biexciton state, respectively. Here, we take into account the repulsive Coulomb interaction between the two electrons V$^{ee}$ and the two holes V$^{hh}$ in the biexciton configuration.\\
Comparing the calculations shown in Fig. 2(c) with the extracted peak positions presented in Fig. 2(b) we observe a very good qualitative agreement. Even the experimentally observed electric field F$_{res}\sim$17kV/cm where 1X$^0$ and 2X$^0$ can be tuned into resonance is in good quantitative agreement with the calculated value when we subtract the critical field F$_{crit}=$4.5 kV/cm which is due to the screening, an effect not included in the calculations.\\
%
\begin{figure}[t]
    \begin{center}
      \includegraphics[width=0.9\columnwidth]{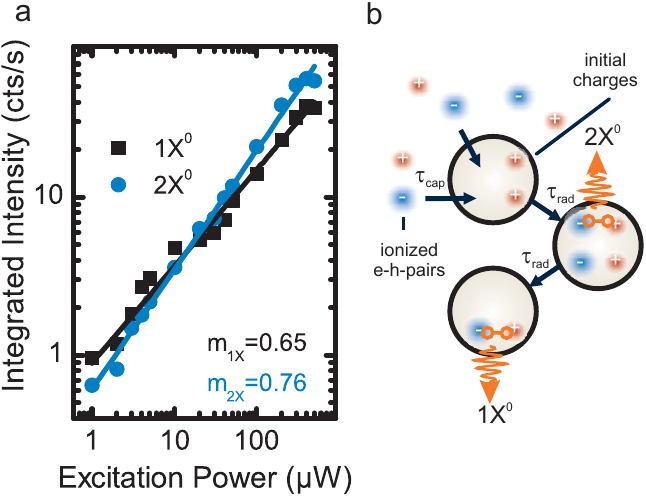}
    \end{center}
    \caption{ (color online) (a) PL intensity of 1X$^0$ and 2X$^0$ as a function of excitation power. (b) Schematic illustration of electron capture in initially charged QDs.}
\end{figure}
%
%
The assignment of different emission lines of single QDs in PL studies is usually done via excitation power dependent measurements (as done for example in Ref. \cite{Finley01}). Here, one typically observes a linear and quadratic behavior of the 1X$^0$ and 2X$^0$ emission lines, respectively. However, as shown in Fig. 3(a) we observe sub-linear behavior for both 1X$^0$ and 2X$^0$ (at $F=7$~kV/cm) with exponents of m$_{1X}=$0.65 and m$_{2X}=$0.76, respectively. We attribute this unusual behavior to non-Poissonian statistic needed to describe the charge carrier capture processes at high electric fields. For electric fields $F>3.3$~kV/cm ionization of optically excited electron-hole pairs occurs and, therefore, single electrons or holes are captured into the QD from the wetting layer rather than bound electron-hole pairs. This further explains the sudden change in PL spectra at electric fields above F$_{crit}$, since either electrons or holes are preferentially captured in initially charged QDs as schematically depicted in Fig. 3(b). In the case of an initially charged QD, one would preferentially capture electrons which leads to a reduction of Coulomb repulsion and, therefore, to a charge neutral dot in the biexciton states. In this case we would not observe linear and quadratic power dependence for exciton and biexciton.\\
Nevertheless, there are indications that our attribution of the emission lines is correct. First of all the energy separation between exciton and biexciton $\Delta=1.3$~meV is typical for InGaAs QDs investigated here and in good agreement with other values reported in the literature \cite{Finley01, Heiss09}. Furthermore, we observe excellent agreement between the measured DC Stark effect and theoretical calculations. Moreover, we performed the same experiments on many different QDs and samples and observed similar results for the exciton biexciton detuning $\Delta$, the power dependence and the Stark shift.\\ 
%
%
In conclusion, we have demonstrated a device consisting of single InGaAs QDs embedded in a lateral electric field geometry. The opposing quantum confined Stark shift for exciton and biexciton enables us  to tune both emission lines in resonance at moderate electric fields $F\sim17$~kV/cm. At resonance the biexciton-exciton radiative cascade is predicted to generate pairs of entangled photons provided time-reordering of the emitted photons \cite{Avron08}. The problem of strongly decreasing PL intensity for increasing electric fields could be solved by exploiting the Purcell effect \cite{Purcell46} in photonic crystal microcavities \cite{Kaniber08, Hofbauer07}.\\
We acknowledge financial support of the Deutsche Forschungsgemeinschaft via the Sonderforschungsbereich 631, Teilprojekt B5 and the German Excellence Initiative via the "Nanosystems Initiative Munich (NIM)".\\

\end{document}